\documentstyle[12pt]{article}

\begin{document}

\begin{titlepage}

\hoffset = -1.3truecm
\voffset = -2truecm

\begin{center}
\bigskip
{\bf \large{Nonlinear Wave Dynamics \\
In Two-Temperature Electron-Positron-Ion Plasma}}
\vskip2truecm
{N. L. SHATASHVILI}\\
\vspace {-0.7em}
{\small \it
{Departamento de Fisica, Universidad de Alcala, E-28871 Alcala de Henares
(Madrid), Spain\\
\vspace {-0.7em}
Plasma Physics Department, Tbilisi State University,
Tbilisi 380028, The Republic of Georgia}}\\
{J. I. JAVAKHISHVILI}\\
\vspace {-0.7em}
{\small \it
{Department of Theoretical Physics, Tbilisi State University, Tbilisi 380028,\\
\vspace {-0.7em}
The Republic of Georgia}}\\
{and}\\
{H. KAYA}\\
\vspace {-0.7em}
{\small \it
{TUBITAK, Marmara Research Center, Research Institute for Basic Sciences,\\
\vspace {-0.7em}
Department of Physics, P.O.Box 21, 41470 Gebze, Kocaeli - Turkey}}\\
\end{center}

\vskip 3.00truecm



\begin{abstract}
\bigskip

The hot electron-positron unmagnetized plasma with small fraction of cold
electron-ion plasma is investigated. The modulational
interactions of the electromagnetic waves and the electron-sound waves are
studied. The possibility of soliton formation is investigated.

\end{abstract}

\end{titlepage}
\centerline{\bf I. Introduction}

During the past few years considereble amount of papers have been devoted to
the analysis of electromagnetic (EM) wave propagation in hot, pure
electron-positron (e-p) plasmas since e-p pairs are thought to be a major
constituent of the plasma emanating both from the pulsars and from the inner
region of the
accretion disks surrounding the central black holes in active galactic nuclei
(AGN) (Michel, 1982; Begelman {\it et al}., 1984). Such a plasma is formed
also in the early universe (Rees, 1983; Tajima and Taniuti, 1980).
Although the relativistically hot e-p pairs form most of astrophysical and
cosmic plasmas, a minority of cold electrons and heavy ions is likely
to be present (Berezhiani and Mahajan, 1995). For instance, outflows of e-p
plamsa from pulsars entering an interstellar cold, low density electron-ion
(e-i) plamsa forms two temperature electron-positron-ion (e-p-i) plasma.
The three-component plasmas - hot e-p plasma with small
fraction of heavy ions - have been studied in the context of pulsar
magnetospheres by Lakhina and Buti (1981) and by Lominadze {\it et al} (1986).
The creation of
stable localized structures of relativistically strong EM radiation in hot
e-p-i plasma have been shown by Berezhiani and Mahajan (1994,1995).

In two temperature e-p-i plasma the interesting phenomena differing
from that of one temperature can exist. It is now believed that strong
monochromatic waves emitted by pulsars are subject to parametric
instabilities even in quite underdence plasmas. In this context in present
paper we consider the propagation of strong EM radiation in a hot e-p
unmagnetized plasma with small fraction of cold e-i plasma.
We show, that the presence of a minority of cold electrons and ions can lead
to the scattering of the pump EM wave into the electron-sound and EM wave;
to the instability of hot e-p plasma against the low frequency (LF)
perturbations.
Hence, in contrast to the case of the pure e-p plasma, in two temperuture
e-p-i plasma the three wave decay
instability may occur. The possibility of the soliton formation due to
the modulational instability of EM wave is also investigated.

\bigskip

\centerline{\bf II. The model and formulations}

Let us assume that the velocity distribution of particles is
locally a relativistic Maxwellian. Then the dynamics of the fluid of species
$\alpha$ ($\alpha=e,p,i$) is contained in the equation (Javakhishvili and
Tsintsadze, 1973):

$${\partial\over\partial x_k}({U_{\alpha}^i}{U_{\alpha k}}{W_{\alpha}})-
{\partial\over \partial x_i}{\it P}_{\alpha}
={1\over c}F^{ik}J_{\alpha k},  \eqno(1)$$
where $U_{\alpha}^i\equiv [\gamma_{\alpha},\ \gamma_{\alpha}{\bf u}_{\alpha}/c]$
is the hydrodynamic four velocity, ${\bf u}_{\alpha}$ is the hydrodynamic
three-velocity
of the hot e-p fluid, $\gamma_{\alpha}=(1-u_{\alpha}^2/c^2)^{-1/2}$ is the
relativistic factor, $J_{\alpha k} $ is the four current, $F^{ik}$ is the
electromagnetic field tensor and $W_{\alpha}$ is the enthalphy per unit
volume: \ \
$W_{\alpha}=(n_{\alpha}/ \gamma_{\alpha})m_{\alpha}c^2 G_{\alpha}
[ m_{\alpha}c^2/ T_{\alpha}] $. \
Here \ $m_{\alpha}$\  and \ $T_{\alpha}$\  are the particle rest mass and
temperature
of species $\alpha$, respectively, \ $n_{\alpha}$\  is the density in the laboratory
frame of the e-p-i fluid and $G_{\alpha}(z_{\alpha})=K_3 (z_{\alpha})/K_2
(z_{\alpha})$, ($z_{\alpha} = m_{\alpha} c^2/T_{\alpha}$), where $K_{\nu}$ are
the modified Bessel functions. For the nonrelativistic temperatures
($T_{\alpha}<<m_{\alpha}c^2$) $G_{\alpha}=1+5T{\alpha}/2m_{\alpha}c^2$ and for
the ultrarelativistic temperatures ($T_{\alpha}>>m_{\alpha}c^2$) $G_{\alpha}=
4T_{\alpha}/m_{\alpha}c^2>>1$. The relativistic pressure in the rest frame is
\ ${\it P}_{\alpha}=(n_{\alpha}/\gamma_{\alpha})T_{\alpha}$.

We assume that hot electron and positron temperatures are equal
and constant while the process of EM wave interaction with given fluid
($G_{\alpha h}=const$). Note that here and below the subscript "$c$" is used
for cold electrons and "$h$" - for hot particles respectively.

>From the set of equations (1) the equation of motion can be written as follows:

$${d_{\alpha}\over dt}({\bf P}_{\alpha}G_{\alpha})+{1\over n_{\alpha}}\nabla
{\it P}_{\alpha}=e_{\alpha}{\bf E}+{e_{\alpha}\over c}({\bf u}_{\alpha}\times
{\bf B}),  \eqno(2)$$
where ${\bf P}_{\alpha}=\gamma_{\alpha}m_{\alpha}{\bf u}_{\alpha}$ is the
hydrodynamic momentum, ${\bf E}$ and ${\bf B}$ are the electric and magnetic
fields and $d_{\alpha}/dt={\partial}/{\partial t}+{\bf u}_{\alpha}\cdot \nabla$
is the comoving derivative. For cold electrons in the eq. (2)
$G_c=1$ and $T_{ec}\equiv T_c=const$ should be assumed.

And for all kinds of species we have the continuity equation:

$${\partial n_{\alpha}\over \partial t}+\nabla (n_{\alpha}
{\bf u}_{\alpha})=0.  \eqno(3)$$

To study the nonlinear propagation of intense EM wave in a relativistically
hot e-p plasma with small fraction of cold e-i plasma we must couple the eq.-s
of motion with the Maxwell equations. In the terms of the potentials defined by:

$${\bf E}=-{1\over c}{\partial{\bf A}\over \partial t}- \nabla \phi ;\ \ \ \ \
{\bf B}= \nabla\times {\bf A},   \eqno(4)$$
they take the form (Coulomb gauge $\nabla\cdot {\bf A}=0$):

$${\partial^2 {\bf A}\over \partial t^2}-c^2 \triangle {\bf A}+c{\partial\over
\partial t}(\nabla \phi)-4\pi c{\bf J}=0,  \eqno(5)$$
and
$$\triangle \phi=-4\pi\rho, \eqno(6)$$
where for the charge and current densities we have respectively:
$$\rho=\sum_{\alpha}e_{\alpha}n_{\alpha} ;\ \ \ \ \ \
{\bf J}=\sum_{\alpha}e_{\alpha}n_{\alpha}{\bf u}_{\alpha} .  \eqno(7)$$

The equilibrium state for hot e-p plasma is characterized with charge neutrality
(with unperturbed number densities of the hot electrons and positrons equal to
$n_{oh}$).  For small fraction of cold e-i plasma the equilibrium state is
characterized also by charge neutrality (with background ion density $n_{oc}$)
and

$$n_{oc}<<n_{oh}  . \eqno(8)$$

Also we assume that: ions are immobile; in equilibrium state hot electrons
and positrons have the same temperatures equal to $T_{oh}$ and

$$T_{oh}>>T_c; \ \ \ \ \ T_{oi}=0 .\eqno(9)$$

Let us analyse the one-dimensional propagation (${\partial\over \partial z}\neq 0,
{\partial\over \partial x}=0, {\partial\over \partial y}=0$) of circularly
polarized EM wave with a mean frequency $\omega_o$ and a mean wave number
$k_o$ along the $z$ axis. Thus

$${\bf A}_{\perp}={1\over 2}({\bf x}+i{\bf y})A(z,t) exp(ik_o z-i\omega_o t)
+c.c. , \eqno(10)$$
where $A(z,t)$ is a slowly verying function of $z$ and $t$ and ${\bf x}$ and
${\bf y}$ are the standard unit vectors. The gauge condition gives us
$A_z=0$. Then the transverse component of eq.-s of motion (2) are
integerated yielding:

$${\bf P}_{\perp\alpha}G_{\alpha}=-{e_{\alpha}\over c}{\bf A}_{\perp},
\eqno(11)$$
where the constant of integration is set equal to zero since particle
hydrodynamic moments are assumed to be zero at the infinity where the field
vanishes.

Now it is necessary to write the equations for longitudinal motion. This motion
is driven by the ponderomotive pressure ($\sim P_{\alpha\perp}^2$) of high
frequency (HF) EM fields and latter doesn't depend on the particle charge sign.
In purely e-p plasma since the effective mass of the electrons and positrons are
equal ($G_e=G_p=G$) the radiation pressure gives equal longitudinal moments
to both the electrons and positrons and effects concentration without producing
the charge separation ($n_e=n_p$ and $\phi=0$) (Berezhiani and Mahajan, 1994;
Kartal {\it et al}, 1995).  But
as it was shown by Berezhiani and Mahajan, (1995) the introduction of small
fraction of heavy ions leads to "symmetry breaking" between hot electrons
and positrons and it becomes possible to have
finite $\phi$. As we will see below, due to the presence of small fraction of
cold e-i plasma in hot e-p plasma the electrostatic potential will be surely
created.

Let us redefine the electron rest mass in equations for hot e-p plasma as:
$$m\to mG_h (T_h)\equiv M .. \eqno(12)$$

Then the eq.(5) for transverse motion can be written as:

$${\partial^2\over\partial t^2}{\bf A}_{\perp}-c^2{\partial^2\over
\partial z^2}{\bf A}_{\perp}+\omega_h^2( n_{eh}
+n_{ph} ) {{\bf A}_{\perp}\over n_{oh}\gamma_h}+
\omega_e^2{n_c\over n_{oc}}{{\bf A}_{\perp}\over \gamma_c}=0,
\eqno(13)$$
where \ $\omega_h^2=4\pi e^2n_{oh}/M$,\ $\omega_e^2=4\pi e^2n_{oc}/m_e$
and $\gamma_h$ and $\gamma_c$ are the
relativistic factors for hot and cold electrons respectively, $\gamma_{\alpha}=
(1+{\bf P}_{\alpha}^2/m_{\alpha}^2c^2)^{1/2}$; \ $n_c,\ n_{eh}$ \  and \ $n_{ph}$
are the cold and hot electron and positron densities respectively.

The equations for longitudinal motion have the form:

$${\partial n_{\alpha}\over \partial t}+{\partial\over \partial z}
\left( {n_{\alpha}P_{z\alpha}\over m_{\alpha}\gamma_{\alpha}}\right)=0,
\eqno(14)$$
$$\left( {\partial\over \partial t}+{P_{z\alpha}\over m_{\alpha}
\gamma_{\alpha}}{\partial\over \partial z}\right) P_{z\alpha}+
{T_{\alpha}\over n_{\alpha}}{\partial\over \partial z}\left(
{n_{\alpha}\over \gamma_{\alpha}}\right) =
-e_{\alpha}{\partial \phi\over \partial z}-{e_{\alpha}^2\over 2m_{\alpha}
\gamma_{\alpha}c^2}{\partial |A|^2\over \partial z}, \eqno(15)$$
where for hot particles $m_{\alpha h}=M$ is assumed and in $P_{z\alpha h}$ the
mass redefinition is performed (see the relation (12)).

In what follows we consider only the weak relativistic case assuming
\ ${\bf P}_\alpha^2/m_\alpha^2c^2<<1$. \ In the presence of two different
temperature electron plasma for LF motion it is possible to satisfy the
condition (Khirseli and Tsintsadze, 1980): \ \ $KV_{Tc}<<\Omega<<KV_{Th}$,
\ \  where $V_{Tc}$ and $V_{Th}$ are the cold and hot electron
thermal velocities respectively and $\Omega^{-1}$ and $K^{-1}$ are the
characteristic time and spacial spreads of the pulse ($\Omega<<\omega_o; \ \
K<<k_o$).

First let's find the equation for LF motion. Under the above mentioned
conditions for hot particles we have:

$$-e_{\alpha}{\partial \phi\over \partial z}={T_h\over n_{\alpha h}}{\partial
\over \partial z}\left( {n_{\alpha h}\over \gamma_h}\right)+{e^2\over 2M
\gamma_h c^2}{\partial |A|^2\over \partial z}. \eqno(16)$$

From eq.-s (16), introducing $\delta n_c=n_c-n_{oc}$ and
$\delta n_{\alpha h}=n_{\alpha h}-n_{oh} \ \ \ (\delta n_c<<n_{oc}; \ \
\delta n_{\alpha h}<<n_{oh})$, we obtain:

$$\delta n_h=\delta n_{ph}+\delta n_{eh}=-n_{oh}{e^2|A|^2\over MT_h c^2}
+ n_{oh}{e^2|A|^2\over M^2c^4}.
\eqno(17)$$

Using equations (6), (7) and (17) we find the relation between $\delta n_{eh}$
and $\delta n_c$ written as:

$$\delta n_{eh}=-{1\over 2}\delta n_c - n_{oh}{e^2|A|^2\over 2MT_h c^2}+
n_{oh}{e^2|A|^2\over 2M^2c^4}. \eqno(18)$$

Using the eq.(18) in the eq.(16) written for electrons finally we obtain:

$${\partial \phi\over \partial z}=-{T_h\over 2e n_{oh}}{\partial\over \partial z}
\delta n_c  . \eqno(19)$$

Thus, as we already mentioned above, the presence of small fraction of cold
e-i plasma in hot e-p plasma gives rise to electrostatic potential.

Substituting the eq.(19) in the eq.(15) written for cold electrons, after simple
algebra, assuming that:

$${T_c\over T_h}<<{1\over 2}{n_{oc}\over n_{oh}} , \eqno(20)$$
one can get the equation for \ $\delta n_c$:

$${\partial^2\over \partial t^2}\delta n_c - c_s^2{\partial^2\over \partial z^2}
\delta n_c = n_{oc}{e^2\over 2m^2c^2}{\partial^2\over \partial z^2}|A|^2,
\eqno(21)$$
where
$$c_s^2=\left( {1\over 2}{n_{oc}\over n_{oh}}{T_h\over m}\right) ^{1/2}$$
is the so-called "electron-sound" velocity. Thus, due to the fact that the
most part of electrons are relativistically hot, and consequently heavy
($G_h\neq 1$)
than the small part of cold electrons, it is possible to induce the
"electron-sound" wave; the exciting ponderomotive force is defined by the
HF pressure on cold electrons.

The eq.(21) is coupled with the following equation for $A$ (for HF wave frequency
satisfying the dispersion relation: $\omega_o^2=k_o^2 c^2 + {2\omega_h^2 } +
\omega_e^2 $):

$$2i \omega_o \left( {\partial\over \partial t}+v_g{\partial\over \partial z}
\right) A + \omega_o v_g^\prime{\partial^2\over \partial z^2}A $$
$$+\omega_h^2 \left( {e^2|A|^2\over M^2c^4}-
{\delta n_h\over n_{oh}}\right)\cdot A
+\omega_e^2\left( {e^2|A|^2\over 2m^2c^4}  -
{\delta n_c\over n_{oc}}\right)\cdot A=0 , \eqno(22)$$
where $v_g$ is the group velocity of HF wave.

The system of equations (21),(22) together with the relation (17) describes
the nonlinear wave dynamics in
a relativistically hot e-p plasma with small fraction of cold e-i plasma. As
we see we have the scattering of EM pump wave into the electron-sound and EM
wave. Note that in the purely e-p plasma the three wave scattering processes
do not exist. The presence of small fraction of cold e-i plasma here is a
reason to have the LF longitudinal waves together with the HF EM waves.
Using this result we can conclude that the radiation emanating both from the
pulsars and AGN entering the
cold low density e-i plasma undergoes the modification due to the scattering
processes.

It is possible to find the stationary solution of the system of eq.-s (17),
(21),(22). We look for the solutions as:

$$A=A(\xi,\tau); \ \ \ \delta n_c=\delta n_c(\xi,\tau); \ \ \
\xi =z-v_g t; \ \ \  t=\tau; \ \ \  {\partial\over \partial \tau}<<v_g{\partial
\over \partial z}. \eqno(23)$$

In the subsonic regime: $v_g<<c_s$, i.e.

$$k_o c^2 \sqrt {m\over M}<<\omega_o \sqrt {{1\over 2}{n_{oc}\over n_{oh}}}V_{Th}
\eqno(24)$$
from the eq.(21) we obtain:

$$\delta n_c=-n_{oc}{c^2\over c_s^2}{e^2|A|^2\over 2m^2c^4}<0 \eqno(25)$$
and substituting the eq.(25) in the eq.(22), if we have the relativistically hot
e-p plasma ($G_h>>1$), we get the Nonlinear Shrodinger Equation (NLSE):

$$2i{\partial\over \partial \tau}A+v_g^\prime{\partial^2\over
\partial \xi^2}A+{\omega_e^2\over \omega_o}{c^2\over c_s^2}
{e^2|A|^2\over 2m^2c^4}\cdot A=0 .\eqno(26)$$

As it is wellknown the eq.(26) has the stationary solution representing
the subsonic soliton of rarification (the total density variation
$\delta n \equiv \delta n_h+\delta n_c <0$).

In the case of the nonrelativistically hot e-p plasma ($G_h\cong 1$) and
$k_o\rightarrow 0$ we aslo get the soliton solution of obtained NLSE.
Under definite conditions it is possible to have the supersonic solitons too.
It should be noted that for the EM waves with $v_g=0$ in pure, hot  e-p
plasma the possibility to have the stable soliton-like structures was found
by Kartal {\it et al,} 1995.

Let's investigate the stability of two-temperature e-p-i plasma. For this
reason we look for $A$ and $\delta n_c$ as:

$$A(z,t)=a(z,t)e^{i\theta (z,t)}; \ \ \
\delta n_c = \delta n_c exp[ikz-i\omega t] + c.c. $$
$$a=a_o + \delta a exp[ikz-i\omega t] + c.c. ; \ \ \
\theta=\theta_o + \delta \theta exp[ikz-i\omega t] + c.c.,  \eqno(27)$$
where $a(z,t)$ and $\theta(z,t)$ are the slowly varying in time and space real
functions and $\delta a<<a_o$, $\delta \theta<<\theta_o$.

Linearizing the system of equations (21), (22) using the relation (17)
we easily obtain the dispersion relation:
\eject

$$(\omega^2-c_s^2k^2)[(w-v_gk)^2-{v_g^{\prime 2}\over 4}k^4]
=\omega_e^2 v_g^{\prime 2}k^4 {e^2a_o^2\over m^2c^4}$$
$$+v_g^{\prime} k^2 (\omega^2-c_s^2k^2){\omega_h^2\over 2\omega_o}
{c^2\over V_{Th}^2}\left( {m^2\over M^2}+{c_s^2\over c^2}\right)
{e^2a_o^2\over m^2c^4}, \eqno(28)$$
which in the limit $M>>m$ (i.e. $G_h>>1$, the relativistically hot e-p plasma)
for the coinciding roots:

$$\omega\simeq c_s k+i\Gamma \simeq v_gk + {v_g^{\prime}k^2\over 2} +
i\Gamma  \eqno(29)$$
gives the relation for increment:

$$\Gamma^2=c_sk{v_g^{\prime}\omega_h^2 \over  V_{Th}^2}
{e^2a_o^2\over m^2c^4} . \eqno(30)$$

Thus, the addition of even very small amount of cold e-i plasma ($n_{oc}\neq 0$,
i.e. $c_s\neq 0$) leads to the instability of hot e-p plasma against the LF
perturbations. Such three wave decay instability doesn't exist in pure e-p plasma.
The present result shoud be useful to understand the character of the pulsar
and AGN radiation.

\bigskip

In conclusion, we have shown that in the hot e-p plasma with small fraction
of cold e-i plasma, it is possible to have the scattering
of EM wave with
relativistically strong amplitude into the longitudinal so-called "electron-
sound" and EM wave. Under the definite conditions the possibility of soliton
solution creation for EM wave is found.

\bigskip

Authors would like to acknowledge N.L.Tsintsadze for his helpful discussions.

\bigskip

The work of J.I.Javakhishvili and N.L.Shatashvili was supported in part
by ISF Long-term Grant KZ3200.

\bigskip
\newpage
\hoffset = -1truecm
\voffset = -2truecm
\hfill
\vskip 1.00truecm

\centerline{\bf References}
\noindent
Begelman, M.C., Blandford, R.D. and Rees, M.D.: 1984, {\it Rev. Mod. Phys.}
{\bf 56}, 255.\\
Berezhiani, V.I. and Mahajan, S.M.: 1994, {\it Phys. Rev. Lett.} {\bf 73}, 1110.\\
Berezhiani, V.I. and Mahajan, S.M.: 1995, {\it Phys. Rev. E}, {\bf 52}, 1968.\\
Javakhishvili, D.I. and Tsintsadze, N.L.: 1973, {\it Zh. Eksp. Teor. Fiz.} {\bf 64},
1314, [1973, {\it Sov. Phys. JETP} {\bf 37}, 666].\\
Kartal, S, Tsintsadze, L.N. and Berezhiani, V.I.: 1995, {\it Phys. Rev. E} ,
{\bf 53}, 4225. \\
Khirseli, E.M. and Tsintsadze, N.L.: 1980, {\it Fizika Plazmy},
{\bf 6}, 1081 [1980, {\it Sov. J. Plasma Phys.} {\bf 6}, 595].\\
Lakhina, G.S. and Buti, B.: 1981, {\it Astrophys. Space Sci.}{\bf 79}, 25.\\
Lominadze, D.C., Machabeli, G.Z., Melikidze, G.I. and Pataraya, A.D.: 1986,
{\it Sov. J. Plasma Phys.} {\bf 12}, 712.\\
Michel, F.C.: 1982, {\it Rev. Mod. Phys.} {\bf 54}, 1.\\
Rees, M.J.: 1983, in G.W. Gibbons, S.W. Hawking and S.Siklos (eds.), {\it The
Very Early Universe}, Cambridge University Press, Cambridge.\\
Tajima, T. and Taniuti, T.: 1990, {\it Phys. Rev. A} {\bf 42}, 3587.

\end{document}